\documentclass[twocolumn,prb,showpacs,floatfix]{revtex4}
\usepackage{graphicx}
\usepackage{amssymb}
\usepackage{bm}

\begin{document}

\title{Collective excitations of hard-core Bosons at
half filling on square and triangular lattices: Development
of roton minima and collapse of roton gap}

\author{Tyler Bryant and Rajiv R.~P.~Singh}
\affiliation{Department of Physics, University of California, Davis,
CA 95616, USA}

\date{\today}

\begin{abstract}
We study ground state properties and excitation spectra for hard-core Bosons
on square and triangular lattices, at half filling, using series expansion 
methods. Nearest-neighbor repulsion between the Bosons leads to the 
development of short-range density order at the antiferromagnetic wavevector,
and simultaneously a roton minima in the density excitation spectra.
On the square-lattice, the model maps on to the well studied XXZ model,
and the roton gap collapses to zero precisely at the Heisenberg
symmetry point, leading to the well known spectra for the Heisenberg
antiferromagnet. On the triangular-lattice, 
the collapse of the roton gap signals the onset of the supersolid phase. Our
results suggest that the transition from the superfluid to the
supersolid phase maybe weakly first order. We also find several features in the
density of states, including two-peaks and a sharp discontinuity, which maybe 
observable in experimental realization of such systems.
\end{abstract}

\maketitle 

\section{Introduction \label{intro}}
A microscopic theory for rotons in the excitation spectra of superfluids 
was first developed by Feynman, where he showed that the roton minima was 
related to a peak in the static structure factor.\cite{feynman54} 
This study has had broad impact in condensed
matter physics ranging from Quantum Hall Effect\cite{gmp} to frustrated
antiferromagnets.\cite{ccl,triangle,triangle-swt} In recent years considerable
interest has also centered on Supersolid phases of matter.\cite{chan} While the
existence of such homogeneous bulk phases in Helium remains
controversial,\cite{supersolid,Anderson05} in case of lattice models such 
phases have been clearly
established. One such example is that of hard-core Bosons hopping on a 
triangular-lattice, where a large enough nearest-neighbor repulsion 
leads to supersolid 
order.\cite{Murthy97,Wessel05,Heidarian05,Melko05,Boninsegni05,Zhao06} 
The nature of the excitation spectra in the superfluid phase and
on approach to the supersolid transition has not been addressed 
for the spin-half model.

Here we use series expansion methods to study the ground state properties
and excitation spectra of hard-core Bosons, at half filling, on square
and triangular lattices, with nearest neighbor repulsion. On the 
square-lattice, the model is equivalent to the antiferromagnetic
XXZ model, and we present the elementary excitation spectra for the XXZ model 
with XY type anisotropy. To our knowledge
this calculation has not been done before. It should be useful for
experimental studies of antiferromagnetic materials with XY anisotropy. 
We set the XY coupling to unity and study the spectra as a function of the 
Ising coupling $J_z$.
For the XY model, the spectra is gapless at $q=0$ (the Goldstone mode of the
superfluid) and has a maximum at the antiferromagnetic wavevector ($\pi,\pi$).
As the Ising coupling is increased a roton minima develops at the
antiferromagnetic wavevector, which goes to zero at the point of Heisenberg 
symmetry ($J_z=1$), as expected for the system with doubled unit cell. 

For the triangular-lattice, the hard-core Boson model maps
onto a ferromagnetic XY model, which is unfrustrated. The nearest-neighbor 
repulsion, on the other hand corresponds to an antiferromagnetic Ising 
coupling, which is frustrated. This model cannot be mapped onto an
{\it antiferromagnetic} XXZ model on the triangular lattice. For this
model, we calculate the equal-time structure factor $S(q)$ as well
as the excitation spectra, $\omega(q)$. Once again, we find that in
the absence of nearest-neighbor repulsion, the excitation spectra is
gapless at $q=0$ and has a maximum at the antiferromagnetic wavevector
(($4\pi/3,0$) and equivalent points). As the repulsion is increased, a
pronounced peak develops in $S(q)$ at these wavevectors and simultaneously a
sharp roton minima develops in the spectra. Series extrapolations suggest
that the roton gap vanishes when the repulsion term ($J_z$) reaches a value 
of $\approx 4.5$. However, we are unable to estimate any critical 
exponents for the vanishing of the gap or for the divergence of the
structure factor. A comparison of our structure factor data with the 
Quantum Monte Carlo data of Wessel and Troyer, leads us to suggest that the 
transition to the supersolid phase maybe weakly first order and occurs for
a value of $J_z$ slightly less than $4.5$.

Our calculations also show a near minimum and flat regions 
in the spectra at the wavevectors 
($\pi,\pi/\sqrt{3}$), which correspond to the midpoint of the faces
of the Brillouin zone.
These are points where the antiferromagnetic
Heisenberg model has a well defined minima.\cite{triangle} In our case the 
dispersion is very flat along some directions and 
a minimum along others. There are several distinguishing features in
the density of states (DOS) of the excitation spectra.
The largest maximum in the DOS is
close to the maximum excitation energy and is not unlike many other antiferromagnets.
But, here, in addition, we get a second maximum in the DOS
from the flat regions in the spectra at the midpoint of the faces of
the Brillouin zone and a sharp drop in the DOS at the roton energy.
It maybe possible to engineer such hard-core Boson systems 
on a triangular-lattice in cold atomic gases. It should, then,
be possible to excite these collective exciations either
optically or by driving the system out of equilibrium. A measurement
of the energies associated with the characteristic features in the density
of states can be used to accurately determine the microscopic
parameters of the system.

\section{Method \label{method}}

The linked-cluster series expansions performed here involve writing 
the Hamiltonian of interest as
\begin{equation}
{\mathcal H}={\mathcal H}_0 + \lambda {\mathcal H}_1
\label{perturb}
\end{equation}
where the eigenstates of ${\mathcal H}_0$ define the basis to be used and ${\mathcal H}_1$ is the 
perturbation to be applied in a linked cluster expansion. Ground state properties are 
then obtained as a power series in $\lambda$ using Raleigh-Schrodinger 
perturbation theory.

Excited state properties are obtained following the procedure outlined in \cite{gelfand00},
in which a similarity transformation is obtained in order to block diagonalize the Hamiltonian 
where the ground state sits in a block by itself and the one-particle states form another block.
\begin{equation}
{\mathcal H}^{\text{eff}}={\mathcal S}^{-1}{\mathcal H}{\mathcal S}
\label{similarity}
\end{equation}
where ${\mathcal H}^{\text{eff}}$ is an effective Hamiltonian for the states which are the 
perturbatively constructed extensions of the single spin-flip states. 
The effective Hamiltonian is then used to obtain a set of transition amplitudes $\sum_{r=0}\lambda^r c_{r,m,n}$ that describe propagation of
the excitation through a distance $(m \hat{x} + n \hat{y})$ for the square lattice and 
$(\frac{1}{2} m\hat{x} +\frac{\sqrt{3}}{2} n \hat{y})$ with $m$ and $n$ both even or both odd for the triangular lattice.

 These transition amplitudes are used to obtain the transition amplitudes for the bulk lattice by summing
over clusters.  Fourier transformation of the bulk transition amplitudes then gives the excitation energy in
momentum space.
\begin{equation}
\Delta(q_x,q_y)=\sum_{r=0}\lambda^r\sum_{m,n}c_{r,m,n}f_{m,n}(q_x,q_y)
\end{equation}
where $f_{m,n}(q_x,q_y)$ is given by the symmetry of the lattice, with
\begin{eqnarray}
f_{m,n}^{\text{sqr}}(q_x,q_y)=\left[cos(m q_x+n q_y)+cos(m q_x-n q_y)\right.\nonumber \\
\left.+cos(n q_x+m q_y)+cos(n q_x-m q_y)\right]/4
\end{eqnarray}
 for the square lattice and 
\begin{eqnarray}
f_{m,n}^{\text{tri}}(q_x,q_y)&=&\left[cos(\frac{m}{2} q_x)cos(\frac{n\sqrt{3}}{2} q_y)\right.\nonumber \\
&+&cos(\frac{\sqrt{3}(m+n)}{4} q_y)cos(\frac{m-3n}{4} q_x) \\
&+&\left.cos(\frac{\sqrt{3}(m-n)}{4} q_y)cos(\frac{m+3n}{4} q_x)\right]/3 \nonumber
\end{eqnarray}
for the triangular lattice.

In order to access values of the expansion parameter $\lambda$ up to and including $\lambda=1$, we use 
standard first order integrated differential approximants \cite{oitmaabook} (IDAs) of the form
\begin{equation}
Q_L(x){df \over dx}+R_M(x)f+S_T(x)=0 
\label{IDA}
\end{equation}
where $Q_L$,$R_M$,$S_T$ are polynomials of degree L,M,and T determined uniquely from the
expansion coefficients.

When gapless modes are present, estimates of the spin-wave velocity are made using the technique
of Singh and Gelfand\cite{singh95}. For small $q=|{\bf q}|$ the spectrum is assumed to have the form $\Delta({\bf q})\sim [A(\lambda)+B(\lambda)q^2]^{1/2}$.
To calculate the spin-wave velocity, we expand $\Delta({\bf q})$ in powers of $q$, $\Delta(q)=C(\lambda) + D(\lambda)q^2 + ...$ and
identify $C=A^{1/2}$ and $D=B/2A^{1/2}$. Thus the series $2C(\lambda)D(\lambda)$ provides an estimate for $B$, which is the square of the 
spin-wave velocity.

\section{Square Lattice \label{square}}

On the square lattice we perform two distinct types of expansions. For $J_z \geq J_\perp$, one can expand
directly in $J_\perp/J_z$ by choosing 
\begin{eqnarray}
{\mathcal H}_0 &=& \sum_{<i,j>} S^z_i S^z_j \nonumber \\
{\mathcal H}_1 &=& \sum_{<i,j>}(S^x_i S^x_j + S^y_i S^y_j)
\end{eqnarray}

In this case $\lambda$ in (\ref{perturb}) is $J_\perp$ (setting $J_z=1$). 
Since ${\mathcal H}_1$ conserves the total $S^z$, one can perform the computation to high order by restricting the full Hilbert space
to the total $S^z$ sector of interest, which in this paper will be restricted to total $S^z=0$ (half filling).  

Series expansion studies of the excitation spectra by Singh et al. \cite{singh95} and subsequently extended
by Zheng et al. \cite{zheng05} have been performed for
$J_z \geq J_\perp$, with expansions involving linked clusters
of up to 11 sites ($\lambda^{10}$) and 15 sites ($\lambda^{14}$) respectively.

\begin{figure}[!htb]
\begin{center}
  \includegraphics[width=\columnwidth,angle=0]{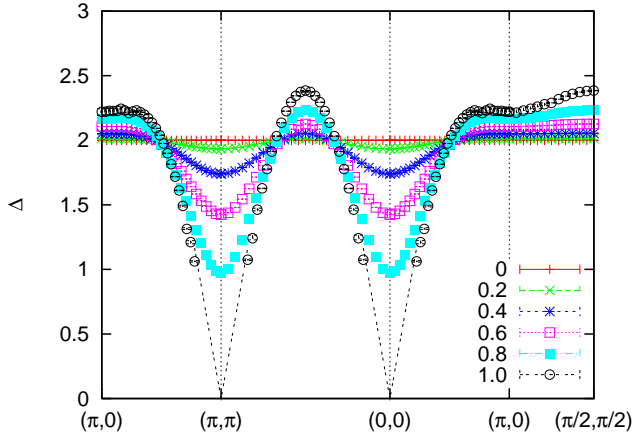}
\end{center}
\vspace{-0.5cm}
\caption{\label{square_bigz_spectrumline}
(Color online) 
The spin-wave dispersion of the XXZ model on the square lattice for various
values of $J_\perp$ ($J_z=1$).  The error bars give an indication of the spread of various IDAs. The
lines around the gapless points for $J_\perp=1$ show the calculated spin-wave velocity. 
}
\end{figure}

Fig. \ref{square_bigz_spectrumline} shows the results of the spin-wave dispersion analysis for $J_\perp$ 
from the dispersionless Ising model $J_\perp = 0$ to the Heisenberg model  $J_\perp = 1$.  One can see the development
of minima at $(0,0)$ and $(\pi,\pi)$ with increasing $J_\perp$, with the gap completely closing at $J_\perp = 1$.
Since IDAs are not accurate near the gapless points, the dotted line shows the estimated spin-wave velocity $v=1.666$ when $J_\perp = 1$.

To obtain the spectra with XY anisotropy ($J_z \leq J_\perp$),
we need to develop a different type of expansion. We consider
the following break up of the Hamiltonian:
(for $J_z \leq J_\perp$) 
\begin{eqnarray}
{\mathcal H}_0 &=& \sum_{<i,j>} S^x_i S^x_j \nonumber \\
{\mathcal H}_1 &=& \sum_{<i,j>} (S^y_i S^y_j + J_z S^z_i S^z_j)
\end{eqnarray} 
where $J_\perp=1$.  Now, a new series is obtained for each value of $J_z$, and the XXZ model is only obtained
upon extrapolation to $\lambda=1$.  In contrast to the first type of expansion, ${\mathcal H}_1$ does
not conserve total $S^z$, and so the entire Hilbert space must be used, limiting the order of computation of the
series to $\lambda^{10}$ (11 sites). 

\begin{figure}[!htb]
\begin{center}
  \includegraphics[width=\columnwidth,angle=0]{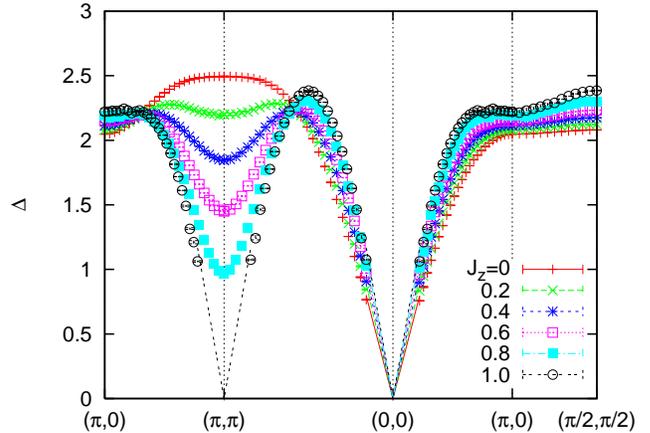}
\end{center}
\vspace{-0.5cm}
\caption{\label{square_smallz_spectrumline}
(Color online) 
The spin-wave dispersion of the XXZ model on the square lattice for various
values of $J_z$ ($J_\perp=1$).  The error bars give an indication of the spread of various IDAs. The
lines around the gapless points show the calculated spin-wave velocities. 
}
\end{figure}

Fig. \ref{square_smallz_spectrumline} shows the results of the spin-wave dispersion analysis for several 
values of $J_z$ from the XY model ($J_z = 0$) to the Heisenberg model ($J_\perp = 1$).
We find that for the pure XY model, there is gapless excitations at $q=0$
(Goldstone modes of the superfluid phase), but there is no roton
minima at the antiferromegnetic wavevector. As $J_z$ is increased,
the spin-wave velocity increases and a clear roton-minima develops
at the antiferromagnetic wavevector. This minima collapses to zero
as the Heisenberg point is approached. In fact, the doubling of
the unit cell implies that for the Heisenberg limit, the spectra at
$q$ and at $q$+($\pi,\pi$) become identical. 
Another point of interest is that along the
direction ($\pi,0$) to ($\pi/2,\pi/2$), which corresponds to the 
antiferromagnetic zone boundary, the dispersion is very flat for
the pure XY model. A weak minimum develops at ($\pi,0$) as the
Heisenberg symmetry point is reached. These results should be useful
in comparing with spectra of two-dimensional antiferromagnets, where
there is significant exchange anisotropy.

\begin{table}[ht]
\caption{Series coefficients for the ground state energy per site $E_0/N$ and $M$}
\begin{tabular}{|c|c|c|}
\hline\hline
n & $E_0/N$ for $J_z$=0 & $M$ for $J_z$=0  \\
\hline 
0 & -5.000000e-01 & 1.250000e-01 \\
2 & -4.166667e-02 & -6.944444e-03 \\
4 & -4.282407e-03 & -2.267072e-03 \\
6 & -1.251190e-03 & -1.141688e-03 \\
8 & -5.538567e-04 & -7.184375e-04 \\
10 & -2.990401e-04 & -5.039687e-04 \\
12 & -1.823004e-04 & -3.784068e-04 \\
14 & -1.015895e-04 & -2.494459e-04 \\
\hline\hline
\end{tabular}
\label{sqr_energyser}
\end{table}

\section{Triangular Lattice \label{triangle}}
There has been much recent interest in the XXZ model on the triangular lattice.
The spin-${1 \over 2}$ XXZ model with ferromagnetic in-plane coupling $J_\perp<0$ and
antiferromagnetic coupling in the $z$ direction $J_z>0$ can be mapped to a hard-core boson model
with nearest neighbor repulsion.
\begin{equation}
  {\mathcal H}_b = -t\sum_{<i,j>} (b^\dag_i b_j + b_i b^\dag_j) + V \sum_{<i,j>}n_i n_j
\label{bosonhamiltonian}
\end{equation}
where $b^\dag_i$ is the bosonic creation operator, $n_i = b^\dag_i b_i$.  The parameters are 
related by $t=-J_\perp/2$ and $V=J_z$.  For the rest of this section, we let $J_\perp=-1$, and
so $V/t = -2J_z/J_\perp = 2J_z$.

We will continue to use the spin language as it is natural for our study.
For $J_z=0$, the ferromagnetic in-plane coupling is unfrustrated. 
As $J_z$ is increased, the competing interaction leads to an emergence
of a supersolid order.

We have performed expansions for the triangular lattice XXZ model of the form
\begin{eqnarray}
{\mathcal H}_0 &=& -\sum_{<i,j>} S^x_i S^x_j \nonumber \\
{\mathcal H}_1 &=& \sum_{<i,j>} (- S^y_i S^y_j + J_z S^z_i S^z_j)
\end{eqnarray} 
where $J_\perp=-1$. Series are obtained for each value of $J_z$, and the XXZ model is obtained
upon extrapolation to $\lambda=1$.

\begin{figure}[!htb]
\begin{center}
  \includegraphics[width=\columnwidth,angle=0]{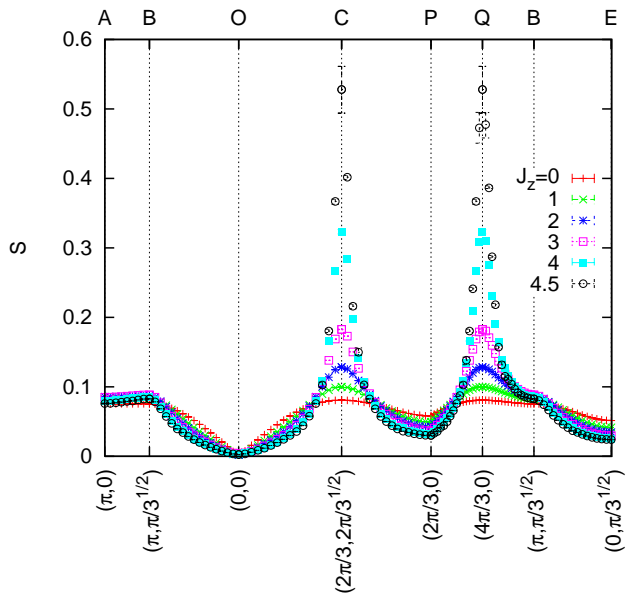}
\end{center}
\vspace{-0.5cm}
\caption{\label{triangle_sfline}
(Color online) 
The static structure factor of the XXZ model on the triangular lattice for various
values of $J_z$ ($J_\perp$=$-1$).  The error bars give an indication of the spread of IDAs.
}
\end{figure}

\begin{figure}[!htb]
\begin{center}
  \includegraphics[width=\columnwidth,angle=0]{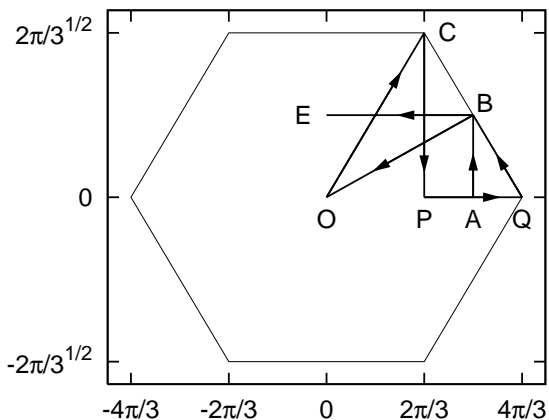}
\end{center}
\vspace{-0.5cm}
\caption{\label{contour}
The hexagonal first Brillouin zone of the triangular lattice and the path ABOCPQBE along which the static structure factor and spin-wave dispersion
have been plotted in Figs. \ref{triangle_sfline} and \ref{triangle_spectrumline}.
}
\end{figure}

\begin{figure}[!htb]
\begin{center}
  \includegraphics[width=\columnwidth,angle=0]{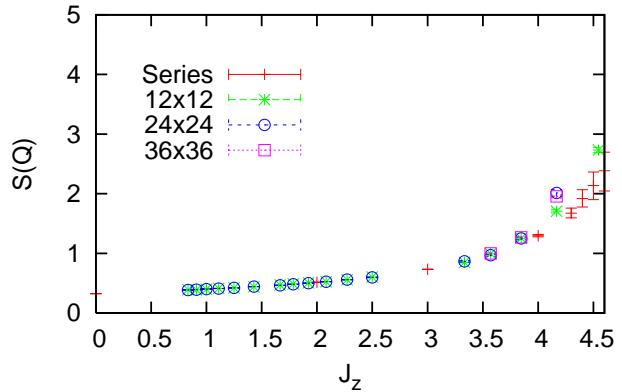}
\end{center}
\vspace{-0.5cm}
\caption{\label{triangle_sfvsjz}
(Color online) 
The static structure factor at ${\bf q}$=$(4\pi/3,0)$ of the XXZ model on the triangular lattice versus $J_z$ ($J_\perp$=$-1$).  
The error bars for the series data give an indication of the spread of IDAs.  Also shown are QMC data
for 12x12,24x24,and 36x36 site clusters from Wessel and Troyer.\cite{Wessel05}
}
\end{figure}

The static structure factor 
\begin{equation}
S({\bf k}) = \sum_{\bf r}e^{i {\bf k}\cdot{\bf r}} \langle S^z_0S^z_{\bf r} \rangle 
\end{equation}
is shown in Fig. \ref{triangle_sfline} along contours shown in
Fig. \ref{contour}.  As $J_z$ increases, a peak forms at wavevector ${\bf q}$=$(4\pi/3,0)$.
A plot of this point is shown in fig. \ref{triangle_sfvsjz} along with QMC data from Wessel and Troyer.\cite{Wessel05}
 
\begin{figure}[!htb]
\begin{center}
  \includegraphics[width=\columnwidth,angle=0]{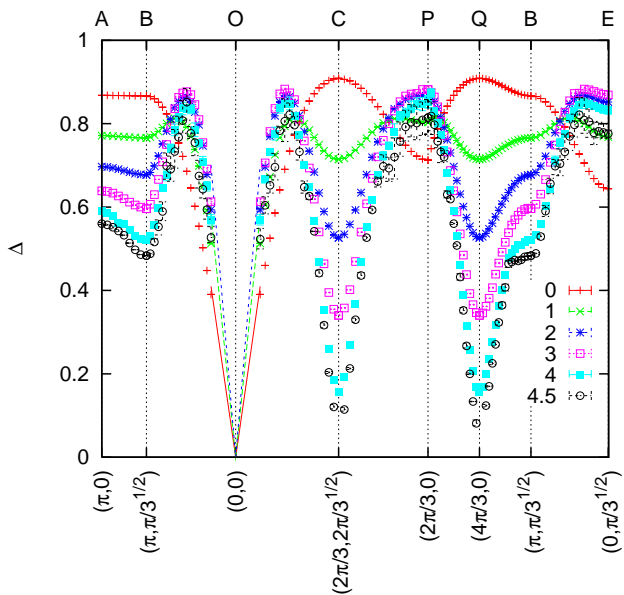}
\end{center}
\vspace{-0.5cm}
\caption{\label{triangle_spectrumline}
(Color online) 
The spin-wave dispersion of the XXZ model on the triangular lattice for various
values of $J_z$ ($J_\perp=-1$).  The error bars give an indication of the spread of IDAs. The
lines around the gapless points show the calculated spin-wave velocities. 
}
\end{figure}

Fig. \ref{triangle_spectrumline} shows the results of the spin-wave dispersion analysis for various
values of $J_z$ ($J_\perp=-1$).  The error bars give an indication of the spread of IDAs. The
lines around the gapless points show the calculated spin-wave velocities. 
One can see the development of minima at ${\bf Q}$ with increasing $J_z$, with the gap completely closing at $J_z \sim 4.5$.
Since IDAs are not accurate near the gapless point ($q=0$), the dotted line shows the estimated spin-wave velocities.

We have been unable to get any consistent estimates for the critical exponents
characterizing the divergence of the antiferromagnetic structure factor and
the vanishing of the roton gap as the supersolid phase is approached.
Furthermore, the comparison with the QMC data of Wessel and Troyer show
that the QMC data begin to show deviations from our series expansion
results before $J_z=4.5$. We believe, this implies that the 
superfluid to supersolid transition is weakly first order. 
Wessel and Troyer estimate the transition to be at 
$J_z\approx 4.3\pm0.2$ ($|t/V|=0.115\pm 0.005$).
Note that the spin-wave theory gives the transition point to be 
at $J_z=2$,\cite{Murthy97} so that quantum fluctuations play a substantial
role here.
Additional QMC studies, should provide further
insight into the nature of the transition.\cite{wessel07}

The calculations also show that near the midpoint of the faces of the Brillouin Zone 
(point B in Fig.~4), the dispersion is 
a minima in the direction perpendicular to the zone face QB and is very flat in
other directions. This behavior 
is reminiscent of the dispersion in the Heisenberg 
antiferromagnet on the traingular lattice where there is
a true minimum at this point.\cite{triangle,triangle-swt}
Note that this behavior is unrelated to any peak in the static structure 
factor and thus, as in case of the Heisenberg model, is
more quantum mechanical in nature. 

\begin{figure}[!htb]
\begin{center}
  \includegraphics[width=\columnwidth,angle=0]{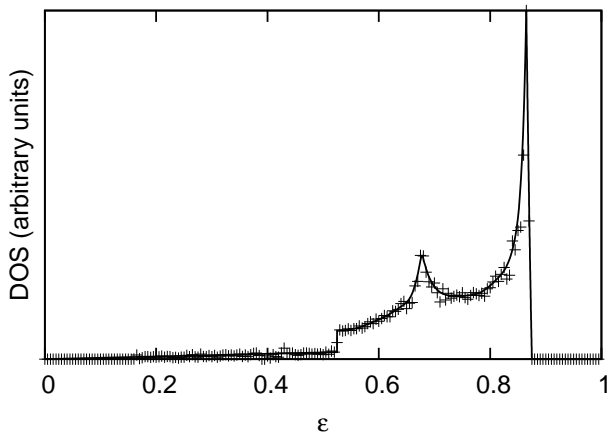}
\end{center}
\vspace{-0.5cm}
\caption{\label{triangle_dos}
The density of states for the XXZ model on the triangular lattice for $J_z=2$ ($J_\perp=-1$).
}
\end{figure}

In Fig.~\ref{triangle_dos}, we show the density of states for the spectra
for $J_z=2$. There are several distinguishing features in the density of
states. First
the largest peak in the density of states occurs close to the highest excitation
energies. This is not unlike what is found in many other antiferromegnets.
However, here, there is a second peak that corresponds to the flat regions in
the spectra near the point B. Finally, at the roton energy there is a sharp
drop in the density of states. The only contributions to the density of states
below the roton gap comes from the Goldstone modes near $q=0$. Since the latter
have very small density of states, there is a discontinuity in the
density of states at the roton energy.

\section{Summary and Conclusions \label{conclusion}}
In this paper, we have studied the excitation spectra of
hard-core Boson models at half-filling on square and triangular
lattices. The calculations show the development of the roton
minima at the antiferromagnetic wavevector, due to nearest-neighbor
repulsion. In accord with Feynman's ideas, the development of the
minima is correlated with the emergence of a sharp peak in
the static structure factor. The case of triangular-lattice is
clearly more interesting as one has a phase transition from a
superfluid to a supersolid phase, where the roton gap goes
to zero. Our series results suggest that the roton-gap vanishes
at $J_z\approx 4.5$. However, there maybe a weakly first order
transition slightly before this $J_z$ value. A more careful finite-size
scaling analysis of the QMC data should provide further insight
into this issue.

Our results of the spectra suggest two peaks in the density of states
and a sharp drop in the density of states at the energy of the roton minima.
If such a hard-core Boson system on a triangular-lattice is
realized in cold-atom experiments, a measurement of the two
peaks in the density of states and the roton minima can be used to determine
independently the hopping parameter $t$ and the nearest-neighbor
repulsion $V$.

\begin{acknowledgements}
This research is supported in part by the National Science Foundation
Grant Number  DMR-0240918. We are greatful to Stefan Wessel for
providing us with the QMC data for the structure factors and to
Marcos Rigol and Stefan Wessel for discussions.
\end{acknowledgements}

\begin{table*}[ht]
\caption{Series coefficients for the magnon dispersion on the square lattice for $J_z=0$ (XY model), nonzero coefficients up to r=9 are listed for compactness (the complete series can be found in Ref.~ \onlinecite{bryantphd})}
\begin{tabular}{|cc|cc|cc|cc|}
\hline\hline
(r,m,n) &  $c_{r,m,n}$ & (r,m,n) &  $c_{r,m,n}$ & (r,m,n) &  $c_{r,m,n}$ & (r,m,n) &  $c_{r,m,n}$ \\
\hline
 (0,0,0) & 2.000000e+00 & (3,2,1) & -7.291667e-02 & (8,3,3) & -1.715283e-03 & (9,5,2) & -1.334928e-03 \\ 
 (2,0,0) & -4.166667e-02 & (5,2,1) & -1.968093e-02 & (4,4,0) & -2.712674e-03 & (8,5,3) & -7.225832e-04 \\ 
 (4,0,0) & -1.023582e-02 & (7,2,1) & -8.897152e-03 & (6,4,0) & -2.980614e-03 & (9,5,4) & -3.993759e-04 \\ 
 (6,0,0) & -5.390283e-03 & (9,2,1) & -4.845897e-03 & (8,4,0) & -1.932294e-03 & (6,6,0) & -1.156309e-04 \\ 
 (8,0,0) & -2.781363e-03 & (4,2,2) & -1.627604e-02 & (5,4,1) & -5.303277e-03 & (8,6,0) & -3.216877e-04 \\ 
 (1,1,0) & -1.000000e+00 & (6,2,2) & -5.758412e-03 & (7,4,1) & -4.614353e-03 & (7,6,1) & -3.803429e-04 \\ 
 (3,1,0) & 4.340278e-02 & (8,2,2) & -2.558526e-03 & (9,4,1) & -3.055177e-03 & (9,6,1) & -6.801940e-04 \\ 
 (5,1,0) & 1.811921e-02 & (3,3,0) & -1.215278e-02 & (6,4,2) & -3.468926e-03 & (8,6,2) & -3.612916e-04 \\ 
 (7,1,0) & 7.679634e-03 & (5,3,0) & -7.265535e-03 & (8,4,2) & -2.946432e-03 & (9,6,3) & -2.662506e-04 \\ 
 (9,1,0) & 4.056254e-03 & (7,3,0) & -3.368594e-03 & (7,4,3) & -1.901715e-03 & (7,7,0) & -2.716735e-05 \\ 
 (2,1,1) & -2.500000e-01 & (9,3,0) & -1.895272e-03 & (9,4,3) & -1.807997e-03 & (9,7,0) & -1.043517e-04 \\ 
 (4,1,1) & -2.314815e-02 & (4,3,1) & -2.170139e-02 & (8,4,4) & -4.516145e-04 & (8,7,1) & -1.032262e-04 \\ 
 (6,1,1) & -5.841368e-03 & (6,3,1) & -1.033207e-02 & (5,5,0) & -5.303277e-04 & (9,7,2) & -1.141074e-04 \\ 
 (8,1,1) & -1.566143e-03 & (8,3,1) & -4.852573e-03 & (7,5,0) & -1.004891e-03 & (8,8,0) & -6.451636e-06 \\ 
 (2,2,0) & -1.250000e-01 & (5,3,2) & -1.060655e-02 & (9,5,0) & -9.122910e-04 & (9,8,1) & -2.852685e-05 \\ 
 (4,2,0) & -3.067130e-02 & (7,3,2) & -6.456544e-03 & (6,5,1) & -1.387570e-03 & (9,9,0) & -1.584825e-06 \\ 
 (6,2,0) & -9.598676e-03 & (9,3,2) & -3.716341e-03 & (8,5,1) & -1.771298e-03 & & \\ 
 (8,2,0) & -3.989385e-03 & (6,3,3) & -2.312617e-03 & (7,5,2) & -1.141029e-03 & & \\ 
\hline \hline
\end{tabular}
\label{sqr_jz0magnon}
\end{table*}

\begin{table*}[ht]
\caption{Series coefficients for the ground state energy per site $E_0/N$ and $M$}
\begin{tabular}{|c|c|c|c|c|c|c|}
\hline\hline
n & $E_0/N$ for $J_z$=0 & $M$ for $J_z$=0  & $E_0/N$ for $J_z$=1 & $M$ for $J_z$=1  & $E_0/N$ for $J_z$=2 & $M$ for $J_z$=2\\
\hline 
0 & -7.500000e-01 & -2.500000e-01  & -7.500000e-01 & -2.500000e-01 & -7.500000e-01 & -2.500000e-01\\
1 & 0.000000e+00 & 0.000000e+00 & & & 0.000000e+00 & 0.000000e+00 \\
2 & -3.750000e-02 & 7.500000e-03  & -1.500000e-01 & 3.000000e-02  & -3.375000e-01 & 6.750000e-02 \\
3 & -7.500000e-03 & 3.000000e-03 & & & 6.750000e-02 & -2.700000e-02 \\
4 & -3.102679e-03 & 1.989902e-03 & -6.428572e-04 & 3.271769e-03  & -3.081696e-02 & 3.263208e-02 \\
5 & -1.557668e-03 & 1.356060e-03 & &  & 4.457109e-02 & -4.706087e-02 \\
6 & -9.211778e-04 & 1.018008e-03 & -1.686432e-03 & 2.253907e-03 & -5.708928e-02 & 7.245005e-02 \\
7 & -5.949646e-04 & 7.975125e-04 & & & 7.181401e-02 & -1.117528e-01 \\
8 & -4.102048e-04 & 6.468307e-04 & -7.027097e-04 & 1.498661e-03 & -1.001587e-01 & 1.839365e-01 \\
9 & -2.965850e-04 & 5.380214e-04 & & & 1.440002e-01 & -3.025679e-01 \\
10 & -2.225228e-04 & 4.565960e-04 & -3.752974e-04 & 1.029273e-03  & -2.164303e-01 & 5.141882e-01 \\
11 & -1.719314e-04 & 3.937734e-04 & & & 3.343757e-01 & -8.849178e-01 \\
12 & -1.360614e-04 & 3.441169e-04 & -2.322484e-04 & 7.779323e-04 & -5.294397e-01 & 1.545967e+00 \\
\hline\hline
\end{tabular}
\label{jzenergyser}
\end{table*}

\begin{table*}[ht]
\caption{Series coefficients for the magnon dispersion on the triangular lattice $J_z=0$, $J_\perp=-1$, nonzero coefficients up to r=9 are listed for compactness (the complete series can be found in Ref.~ \onlinecite{bryantphd})}
\begin{tabular}{|cc|cc|cc|cc|}
\hline\hline
(r,m,n) &  $c_{r,m,n}$ & (r,m,n) &  $c_{r,m,n}$ & (r,m,n) &  $c_{r,m,n}$ & (r,m,n) &  $c_{r,m,n}$ \\
\hline
(0,0,0) & 3.000000e+00 & (6,5,1) & -6.186994e-03 & (5,8,0) & -2.318836e-03 & (9,10,6) & -3.135408e-05 \\ 
 (2,0,0) & -3.125000e-02 & (7,5,1) & -4.204807e-03 & (6,8,0) & -2.496675e-03 & (9,10,8) & -4.493212e-07 \\ 
 (3,0,0) & 6.927083e-03 & (8,5,1) & -3.031857e-03 & (7,8,0) & -2.214348e-03 & (6,11,1) & -9.073618e-05 \\ 
 (4,0,0) & -5.786823e-03 & (9,5,1) & -2.274746e-03 & (8,8,0) & -1.849982e-03 & (7,11,1) & -2.637766e-04 \\ 
 (5,0,0) & -2.071746e-03 & (4,5,3) & -3.466797e-03 & (9,8,0) & -1.528161e-03 & (8,11,1) & -3.828143e-04 \\ 
 (6,0,0) & -2.263644e-03 & (5,5,3) & -4.350420e-03 & (5,8,2) & -1.098718e-03 & (9,11,1) & -4.387479e-04 \\ 
 (7,0,0) & -1.418733e-03 & (6,5,3) & -3.723124e-03 & (6,8,2) & -1.554950e-03 & (7,11,3) & -7.642054e-05 \\ 
 (8,0,0) & -1.158308e-03 & (7,5,3) & -2.959033e-03 & (7,8,2) & -1.579331e-03 & (8,11,3) & -1.651312e-04 \\ 
 (9,0,0) & -8.857566e-04 & (8,5,3) & -2.305863e-03 & (8,8,2) & -1.427534e-03 & (9,11,3) & -2.316277e-04 \\ 
 (1,2,0) & -1.500000e+00 & (9,5,3) & -1.817374e-03 & (9,8,2) & -1.241335e-03 & (8,11,5) & -1.824976e-05 \\ 
 (2,2,0) & -2.500000e-01 & (3,6,0) & -7.265625e-03 & (6,8,4) & -2.268405e-04 & (9,11,5) & -4.925111e-05 \\ 
 (3,2,0) & 2.236979e-02 & (4,6,0) & -1.113487e-02 & (7,8,4) & -4.458355e-04 & (9,11,7) & -1.797285e-06 \\ 
 (4,2,0) & -3.470238e-03 & (5,6,0) & -7.839022e-03 & (8,8,4) & -5.558869e-04 & (6,12,0) & -1.512270e-05 \\ 
 (5,2,0) & 6.223272e-03 & (6,6,0) & -5.440344e-03 & (9,8,4) & -5.861054e-04 & (7,12,0) & -9.486853e-05 \\ 
 (6,2,0) & 2.258837e-03 & (7,6,0) & -3.834575e-03 & (7,8,6) & -1.528411e-05 & (8,12,0) & -1.917140e-04 \\ 
 (7,2,0) & 2.729869e-03 & (8,6,0) & -2.794337e-03 & (8,8,6) & -6.113630e-05 & (9,12,0) & -2.589374e-04 \\ 
 (8,2,0) & 1.793769e-03 & (9,6,0) & -2.110041e-03 & (9,8,6) & -1.121505e-04 & (7,12,2) & -4.585233e-05 \\ 
 (9,2,0) & 1.431100e-03 & (4,6,2) & -5.200195e-03 & (5,9,1) & -5.493588e-04 & (8,12,2) & -1.209317e-04 \\ 
 (2,3,1) & -1.875000e-01 & (5,6,2) & -5.161886e-03 & (6,9,1) & -1.094328e-03 & (9,12,2) & -1.836224e-04 \\ 
 (3,3,1) & -5.677083e-02 & (6,6,2) & -4.187961e-03 & (7,9,1) & -1.233659e-03 & (8,12,4) & -2.281220e-05 \\ 
 (4,3,1) & -2.743217e-02 & (7,6,2) & -3.218811e-03 & (8,9,1) & -1.182811e-03 & (9,12,4) & -5.685542e-05 \\ 
 (5,3,1) & -1.275959e-02 & (8,6,2) & -2.454043e-03 & (9,9,1) & -1.066119e-03 & (9,12,6) & -4.193665e-06 \\ 
 (6,3,1) & -8.214468e-03 & (9,6,2) & -1.905072e-03 & (6,9,3) & -3.024539e-04 & (7,13,1) & -1.528411e-05 \\ 
 (7,3,1) & -4.959005e-03 & (5,6,4) & -5.493588e-04 & (7,9,3) & -5.232866e-04 & (8,13,1) & -6.113630e-05 \\ 
 (8,3,1) & -3.456161e-03 & (6,6,4) & -1.094328e-03 & (8,9,3) & -6.255503e-04 & (9,13,1) & -1.121505e-04 \\ 
 (9,3,1) & -2.443109e-03 & (7,6,4) & -1.233659e-03 & (9,9,3) & -6.426095e-04 & (8,13,3) & -1.824976e-05 \\ 
 (2,4,0) & -9.375000e-02 & (8,6,4) & -1.182811e-03 & (7,9,5) & -4.585233e-05 & (9,13,3) & -4.925111e-05 \\ 
 (3,4,0) & -4.916667e-02 & (9,6,4) & -1.066119e-03 & (8,9,5) & -1.209317e-04 & (9,13,5) & -6.290497e-06 \\ 
 (4,4,0) & -2.605934e-02 & (4,7,1) & -3.466797e-03 & (9,9,5) & -1.836224e-04 & (7,14,0) & -2.183444e-06 \\ 
 (5,4,0) & -1.289340e-02 & (5,7,1) & -4.350420e-03 & (8,9,7) & -2.607109e-06 & (8,14,0) & -1.878307e-05 \\ 
 (6,4,0) & -8.269582e-03 & (6,7,1) & -3.723124e-03 & (9,9,7) & -1.376709e-05 & (9,14,0) & -4.937808e-05 \\ 
 (7,4,0) & -5.280757e-03 & (7,7,1) & -2.959033e-03 & (5,10,0) & -1.098718e-04 & (8,14,2) & -9.124881e-06 \\ 
 (8,4,0) & -3.753955e-03 & (8,7,1) & -2.305863e-03 & (6,10,0) & -4.726374e-04 & (9,14,2) & -3.135408e-05 \\ 
 (9,4,0) & -2.735539e-03 & (9,7,1) & -1.817374e-03 & (7,10,0) & -7.110879e-04 & (9,14,4) & -6.290497e-06 \\ 
 (3,4,2) & -2.179687e-02 & (5,7,3) & -1.098718e-03 & (8,10,0) & -7.825389e-04 & (8,15,1) & -2.607109e-06 \\ 
 (4,4,2) & -1.596136e-02 & (6,7,3) & -1.554950e-03 & (9,10,0) & -7.669912e-04 & (9,15,1) & -1.376709e-05 \\ 
 (5,4,2) & -9.470639e-03 & (7,7,3) & -1.579331e-03 & (6,10,2) & -2.268405e-04 & (9,15,3) & -4.193665e-06 \\ 
 (6,4,2) & -6.186994e-03 & (8,7,3) & -1.427534e-03 & (7,10,2) & -4.458355e-04 & (8,16,0) & -3.258886e-07 \\ 
 (7,4,2) & -4.204807e-03 & (9,7,3) & -1.241335e-03 & (8,10,2) & -5.558869e-04 & (9,16,0) & -3.685726e-06 \\ 
 (8,4,2) & -3.031857e-03 & (6,7,5) & -9.073618e-05 & (9,10,2) & -5.861054e-04 & (9,16,2) & -1.797285e-06 \\ 
 (9,4,2) & -2.274746e-03 & (7,7,5) & -2.637766e-04 & (7,10,4) & -7.642054e-05 & (9,17,1) & -4.493212e-07 \\ 
 (3,5,1) & -2.179687e-02 & (8,7,5) & -3.828143e-04 & (8,10,4) & -1.651312e-04 & (9,18,0) & -4.992458e-08 \\ 
 (4,5,1) & -1.596136e-02 & (9,7,5) & -4.387479e-04 & (9,10,4) & -2.316277e-04 & & \\ 
 (5,5,1) & -9.470639e-03 & (4,8,0) & -8.666992e-04 & (8,10,6) & -9.124881e-06 & & \\ 
\hline \hline
\end{tabular}
\label{jz0magnon}
\end{table*}

\begin{table*}[ht]
\caption{Series coefficients for the magnon dispersion on the triangular lattice $J_z=1$, $J_\perp=-1$, nonzero coefficients up to r=9 are listed for compactness (the complete series can be found in Ref.~ \onlinecite{bryantphd})}
\begin{tabular}{|cc|cc|cc|cc|}
\hline\hline
(r,m,n) &  $c_{r,m,n}$ & (r,m,n) &  $c_{r,m,n}$ & (r,m,n) &  $c_{r,m,n}$ & (r,m,n) &  $c_{r,m,n}$ \\
\hline
 (0,0,0) & 3.000000e+00 & (8,4,2) & 1.641818e-02 & (8,7,3) & -3.898852e-03 & (6,10,2) & -1.174079e-03 \\ 
 (2,0,0) & -1.250000e-01 & (4,5,1) & -7.072545e-02 & (6,7,5) & -4.696316e-04 & (8,10,2) & -2.957178e-03 \\ 
 (4,0,0) & 8.773202e-02 & (6,5,1) & -2.266297e-02 & (8,7,5) & -2.220857e-03 & (8,10,4) & -1.051498e-03 \\ 
 (6,0,0) & -3.747008e-02 & (8,5,1) & 1.641818e-02 & (4,8,0) & -3.632812e-03 & (8,10,6) & -5.518909e-05 \\ 
 (8,0,0) & 2.658737e-02 & (4,5,3) & -1.453125e-02 & (6,8,0) & -1.296471e-02 & (6,11,1) & -4.696316e-04 \\ 
 (2,2,0) & -1.000000e-00 & (6,5,3) & -1.709052e-02 & (8,8,0) & -2.634458e-03 & (8,11,1) & -2.220857e-03 \\ 
 (4,2,0) & 3.067262e-01 & (8,5,3) & -1.306500e-05 & (6,8,2) & -8.548979e-03 & (8,11,3) & -1.051498e-03 \\ 
 (6,2,0) & -1.478629e-01 & (4,6,0) & -4.674479e-02 & (8,8,2) & -3.898852e-03 & (8,11,5) & -1.103782e-04 \\ 
 (8,2,0) & 1.192248e-01 & (6,6,0) & -2.107435e-02 & (6,8,4) & -1.174079e-03 & (6,12,0) & -7.827194e-05 \\ 
 (2,3,1) & -7.500000e-01 & (8,6,0) & 8.083746e-03 & (8,8,4) & -2.957178e-03 & (8,12,0) & -1.178397e-03 \\ 
 (4,3,1) & 8.683532e-02 & (4,6,2) & -2.179687e-02 & (8,8,6) & -3.787294e-04 & (8,12,2) & -7.614846e-04 \\ 
 (6,3,1) & -7.198726e-02 & (6,6,2) & -1.775987e-02 & (6,9,1) & -5.919357e-03 & (8,12,4) & -1.379727e-04 \\ 
 (8,3,1) & 6.383529e-02 & (8,6,2) & 1.136255e-03 & (8,9,1) & -4.075885e-03 & (8,13,1) & -3.787294e-04 \\ 
 (2,4,0) & -3.750000e-01 & (6,6,4) & -5.919357e-03 & (6,9,3) & -1.565439e-03 & (8,13,3) & -1.103782e-04 \\ 
 (4,4,0) & -2.357440e-02 & (8,6,4) & -4.075885e-03 & (8,9,3) & -3.182356e-03 & (8,14,0) & -1.146580e-04 \\ 
 (6,4,0) & -4.667627e-02 & (4,7,1) & -1.453125e-02 & (8,9,5) & -7.614846e-04 & (8,14,2) & -5.518909e-05 \\ 
 (8,4,0) & 4.476721e-02 & (6,7,1) & -1.709052e-02 & (8,9,7) & -1.576831e-05 & (8,15,1) & -1.576831e-05 \\ 
 (4,4,2) & -7.072545e-02 & (8,7,1) & -1.306500e-05 & (6,10,0) & -2.499303e-03 & (8,16,0) & -1.971039e-06 \\ 
 (6,4,2) & -2.266297e-02 & (6,7,3) & -8.548979e-03 & (8,10,0) & -3.646508e-03 & & \\ 
\hline \hline
\end{tabular}
\label{jz1magnon}
\end{table*}

\begin{table*}[ht]
\caption{Series coefficients for the magnon dispersion on the triangular lattice $J_z=2$, $J_\perp=-1$, nonzero coefficients up to r=9 are listed for compactness (the complete series can be found in Ref.~ \onlinecite{bryantphd})}
\begin{tabular}{|cc|cc|cc|cc|}
\hline\hline
(r,m,n) &  $c_{r,m,n}$ & (r,m,n) &  $c_{r,m,n}$ & (r,m,n) &  $c_{r,m,n}$ & (r,m,n) &  $c_{r,m,n}$ \\
\hline
 (0,0,0) & 3.000000e+00 & (6,5,1) & -5.113269e-01 & (5,8,0) & 1.114660e-01 & (9,10,6) & 3.094665e-02 \\
 (2,0,0) & -2.812500e-01 & (7,5,1) & 3.541550e-01 & (6,8,0) & -2.789923e-01 & (9,10,8) & 4.332352e-04 \\
 (3,0,0) & -6.234375e-02 & (8,5,1) & -5.927875e-01 & (7,8,0) & 3.889189e-01 & (6,11,1) & -1.020493e-02 \\
 (4,0,0) & 3.427127e-01 & (9,5,1) & 1.818953e+00 & (8,8,0) & -5.937472e-01 & (7,11,1) & 5.778696e-02 \\
 (5,0,0) & 1.248868e-01 & (4,5,3) & -9.659180e-02 & (9,8,0) & 1.024914e+00 & (8,11,1) & -1.791293e-01 \\
 (6,0,0) & -3.138557e-01 & (5,5,3) & 2.102338e-01 & (5,8,2) & 5.215530e-02 & (9,11,1) & 3.824414e-01 \\
 (7,0,0) & -2.065582e-01 & (6,5,3) & -3.793451e-01 & (6,8,2) & -1.818856e-01 & (7,11,3) & 1.646341e-02 \\
 (8,0,0) & 2.679692e-01 & (7,5,3) & 4.434905e-01 & (7,8,2) & 3.088576e-01 & (8,11,3) & -8.048695e-02 \\
 (9,0,0) & 6.130976e-01 & (8,5,3) & -6.470161e-01 & (8,8,2) & -5.178979e-01 & (9,11,3) & 2.165711e-01 \\
 (1,2,0) & 1.500000e+00 & (9,5,3) & 1.205017e+00 & (9,8,2) & 8.731811e-01 & (8,11,5) & -8.618625e-03 \\
 (2,2,0) & -2.250000e+00 & (3,6,0) & 6.539063e-02 & (6,8,4) & -2.551233e-02 & (9,11,5) & 4.893766e-02 \\
 (3,2,0) & -2.013281e-01 & (4,6,0) & -3.105654e-01 & (7,8,4) & 9.867811e-02 & (9,11,7) & 1.732941e-03 \\
 (4,2,0) & 1.349036e+00 & (5,6,0) & 3.819521e-01 & (8,8,4) & -2.507655e-01 & (6,12,0) & -1.700822e-03 \\
 (5,2,0) & 3.245835e-01 & (6,6,0) & -4.794666e-01 & (9,8,4) & 4.867085e-01 & (7,12,0) & 2.057252e-02 \\
 (6,2,0) & -9.772865e-01 & (7,6,0) & 4.227344e-01 & (7,8,6) & 3.292682e-03 & (8,12,0) & -9.194406e-02 \\
 (7,2,0) & -1.324972e+00 & (8,6,0) & -6.511353e-01 & (8,8,6) & -2.934241e-02 & (9,12,0) & 2.400559e-01 \\
 (8,2,0) & 1.736288e+00 & (9,6,0) & 1.561357e+00 & (9,8,6) & 1.091019e-01 & (7,12,2) & 9.878045e-03 \\
 (9,2,0) & 1.998149e+00 & (4,6,2) & -1.448877e-01 & (5,9,1) & 2.607765e-02 & (8,12,2) & -5.858223e-02 \\
 (2,3,1) & -1.687500e+00 & (5,6,2) & 2.496909e-01 & (6,9,1) & -1.264872e-01 & (9,12,2) & 1.748291e-01 \\
 (3,3,1) & 5.109375e-01 & (6,6,2) & -4.028414e-01 & (7,9,1) & 2.546730e-01 & (8,12,4) & -1.077328e-02 \\
 (4,3,1) & 1.438694e-01 & (7,6,2) & 4.542205e-01 & (8,9,1) & -4.585226e-01 & (9,12,4) & 5.664036e-02 \\
 (5,3,1) & 4.631412e-01 & (8,6,2) & -6.635572e-01 & (9,9,1) & 7.791699e-01 & (9,12,6) & 4.043528e-03 \\
 (6,3,1) & -7.686440e-01 & (9,6,2) & 1.278006e+00 & (6,9,3) & -3.401644e-02 & (7,13,1) & 3.292682e-03 \\
 (7,3,1) & -2.673657e-01 & (5,6,4) & 2.607765e-02 & (7,9,3) & 1.163420e-01 & (8,13,1) & -2.934241e-02 \\
 (8,3,1) & 2.424913e-01 & (6,6,4) & -1.264872e-01 & (8,9,3) & -2.770943e-01 & (9,13,1) & 1.091019e-01 \\
 (9,3,1) & 2.292894e+00 & (7,6,4) & 2.546730e-01 & (9,9,3) & 5.242482e-01 & (8,13,3) & -8.618625e-03 \\
 (2,4,0) & -8.437500e-01 & (8,6,4) & -4.585226e-01 & (7,9,5) & 9.878045e-03 & (9,13,3) & 4.893766e-02 \\
 (3,4,0) & 4.425000e-01 & (9,6,4) & 7.791699e-01 & (8,9,5) & -5.858223e-02 & (9,13,5) & 6.065292e-03 \\
 (4,4,0) & -3.406189e-01 & (4,7,1) & -9.659180e-02 & (9,9,5) & 1.748291e-01 & (7,14,0) & 4.703831e-04 \\
 (5,4,0) & 5.419257e-01 & (5,7,1) & 2.102338e-01 & (8,9,7) & -1.231232e-03 & (8,14,0) & -8.933341e-03 \\
 (6,4,0) & -6.661367e-01 & (6,7,1) & -3.793451e-01 & (9,9,7) & 1.347384e-02 & (9,14,0) & 4.854225e-02 \\
 (7,4,0) & 3.675761e-02 & (7,7,1) & 4.434905e-01 & (5,10,0) & 5.215530e-03 & (8,14,2) & -4.309313e-03 \\
 (8,4,0) & -2.016564e-01 & (8,7,1) & -6.470161e-01 & (6,10,0) & -5.380211e-02 & (9,14,2) & 3.094665e-02 \\
 (9,4,0) & 2.293150e+00 & (9,7,1) & 1.205017e+00 & (7,10,0) & 1.541458e-01 & (9,14,4) & 6.065292e-03 \\
 (3,4,2) & 1.961719e-01 & (5,7,3) & 5.215530e-02 & (8,10,0) & -3.358010e-01 & (8,15,1) & -1.231232e-03 \\
 (4,4,2) & -4.619167e-01 & (6,7,3) & -1.818856e-01 & (9,10,0) & 6.055844e-01 & (9,15,1) & 1.347384e-02 \\
 (5,4,2) & 4.570889e-01 & (7,7,3) & 3.088576e-01 & (6,10,2) & -2.551233e-02 & (9,15,3) & 4.043528e-03 \\
 (6,4,2) & -5.113269e-01 & (8,7,3) & -5.178979e-01 & (7,10,2) & 9.867811e-02 & (8,16,0) & -1.539040e-04 \\
 (7,4,2) & 3.541550e-01 & (9,7,3) & 8.731811e-01 & (8,10,2) & -2.507655e-01 & (9,16,0) & 3.577909e-03 \\
 (8,4,2) & -5.927875e-01 & (6,7,5) & -1.020493e-02 & (9,10,2) & 4.867085e-01 & (9,16,2) & 1.732941e-03 \\
 (9,4,2) & 1.818953e+00 & (7,7,5) & 5.778696e-02 & (7,10,4) & 1.646341e-02 & (9,17,1) & 4.332352e-04 \\
 (3,5,1) & 1.961719e-01 & (8,7,5) & -1.791293e-01 & (8,10,4) & -8.048695e-02 & (9,18,0) & 4.813724e-05 \\
 (4,5,1) & -4.619167e-01 & (9,7,5) & 3.824414e-01 & (9,10,4) & 2.165711e-01 & & \\
 (5,5,1) & 4.570889e-01 & (4,8,0) & -2.414795e-02 & (8,10,6) & -4.309313e-03 & & \\
\hline \hline
\end{tabular}
\label{jz47magnon}
\end{table*}

\end{document}